\documentclass[a4paper]{jpconf}
\usepackage{graphicx}
\begin{document}
\title{Strangeness production and long-range correlations in pp collisions in string fusion approach}

\author{Vladimir Kovalenko, Vladimir Vechernin}

\address{Saint Petersburg State University, Russia}

\ead{v.kovalenko@spbu.ru}

\begin{abstract}
The effects of string fusion on the correlations in strange particles production in proton-proton collisions at high energy are studied in the framework of a Monte Carlo string-parton model. The model is based on the strings formation in elementary dipole-dipole collisions. The hardness of the elementary interaction is defined by a transverse size of the colliding dipoles.
The interaction between strings is realized in the accordance with the string fusion model prescriptions by the introduction of the lattice in the impact parameter plane and taking into account the finite rapidity length of strings. The particles species differentiation is implemented according to Schwinger mechanism. The parameters of the model are fixed with the experimental data on total inelastic cross section and charged multiplicity. 
In the framework of the model the long-range correlation functions with an accounting of strangeness have been studied. A new intensive event-by-event observable has been proposed, which characterizes the fraction of strange particles in the event. The predictions on the correlations between strangeness, multiplicity and mean transverse momentum are obtained for pp collisions at 7~TeV.
\end{abstract}

\section{Introduction}

Long-range (forward-backward) correlation studies between observables in two separated rapidity windows
are considered \cite{b_ALICE} as a tool for investigation
of the initial stages of the hadronic and nuclear collisions.
%in case of ultra-relativistic heavy ion collisions
%preceding the creation of 
%a hot and dense medium.
Due to the non-perturbative nature of multiparticle production in a soft region, one has to apply the  semiphenomenological approaches, such as 
the model of quark-gluon string formation.
At high energies, because of multiple partonic interactions, the formation of several pairs of strings becomes possible. The interaction between the strings could be observed as collective phenomena in pp collisions. Particularly, according to the string fusion model \cite{SFa,SFb,SFc}, the multiplicity, transverse momentum of produced particles, as well as the yield of strangeness are modified compared to the ones of independent strings. This results in a non-trivial correlation between strangeness, multiplicity and transverse momentum.

Numerically, the long-range correlations are studied in terms of correlation functions and correlation coefficients.
A correlation function by definition is
the mean value $\langle B \rangle_F$ of variable B
in the backward window as a function of
another variable F in the forward rapidity
window:
%\begin{equation}
$f_{B\mathrm{-}F}(F)={\langle B\rangle}_F.$
%\end{equation}
The correlation coefficient is a slope
of correlation function:
%\begin{equation} \label{defcorr}
${b_{B\mathrm{-}F}=\frac{df_{B\mathrm{-}F}(F)}{dF}|_{{F}=<{F}>}.}$
%\end{equation}
The role of the observables, $B$ and $F$, in backward and
 forward rapidity windows can play multiplicity ($n$)
 or mean event transverse momentum ($p_t$=$\frac{1}{n}\sum\limits_{i=1}^{n} {p_t}_i$) \cite{SFc,Vest1,Vest2}.

In the present paper, we introduce an additional variable for forward-backward correlation studies, which is defined as an event fraction of strange particles in a given rapidity interval~--~$S$. 
We study correlation functions in the framework of the Monte Carlo model \cite{MC_model,MC_model2}, that incorporates string collectivity effects in the form of string fusion \cite{SFa,SFb,SFc}, and provide the predictions for $pp$ collisions at 7~TeV.

\section{Monte Carlo model}

The Monte Carlo model \cite{MC_model,MC_model2} is based on the partonic picture of nucleon interaction. It preserves the energy and angular momentum conservation in the nucleon initial state 
and uses the dipole approach \cite{hardness} for description of elementary partonic collisions.
Multiplicity and transverse momentum are obtained in the approach of color strings, stretched between projectile and
target partons.
The interaction of strings is realized in accordance with the
string fusion model prescriptions \cite{SFa,SFb,SFc}.
Namely, the mean multiplicity $\mu$ and
 the mean transverse momentum $p_{T}$ of the particles produced from a cluster of $k$ overlapping strings are related to those ($\mu_1, {p_T}_1$) from a single string, as follows: 
$\mu= \sqrt{k} \mu_{1}, p_{T}=\sqrt[4]{k} {p_T}_{1}.$ 
%For realization of the string fusion prescription, the discrete approach has been used, in which a lattice with the cell area equal to the string transverse area $\pi r_{\mathrm{str}}^2$ is introduced. The strings are fused if their transverse position centres belong to the same cell.
For the multiplicity from one string (or a cluster of fused strings) we used Poisson
distribution, with Gaussian transverse momentum spectra of produced particles.
%\subsection{Hardness of the process}

In order to provide the possibility of a direct comparison with experimental data, including the correct description of the transverse momentum spectra, the
MC model \cite{MC_model,MC_model2} has been extended \cite{hardness} by taking into account the hardness of elementary collision. For this purpose the mechanism similar to the one in DIPSY event generator has been incorporated in our model
with the string fusion.
It was assumed that the hardness an elementary collision is inversely proportional to the
transverse size of the interacting dipoles:
$
{d}_i=|\vec r_1 - \vec r_2 |,  d_i'=|\vec {r_1}' - \vec {r_2}'|.
$
Therefore, the mean transverse momentum of particles produced by a single string has the contributions from both edges of the string plus the additional constant term $p_0$, corresponding to the intrinsic string transverse momentum:
${{p_{T}}_{1}}^2=\frac{1}{d_i^2}+\frac{1}{{d_i'}^2} + p_0^2$.
Accordingly, in the version with string fusion, the transverse momentum of a cluster of strings: 
$p_T^4 = \sum_{i=1}^k {{p_{T}}_{1_i}}^4,$
where ${{p_{T}}_{1_i}}^2=\frac{1}{d_i^2}+\frac{1}{{d_i'}^2} + p_0^2$.

Parameters of the model are constrained from the data on total
inelastic cross-section and multiplicity \cite{MC_model3}. We used the string radius $r_\mathrm{str}=0.2$~fm, and the intrinsic string transverse momentum  ${p_0 = 0.2 \mathrm{~GeV}/c}$, which provides a reasonable description of the transverse momentum distribution in $pp$ collisions at the LHC energies.

The particles species differentiation in this model is implemented in the framework of Schwinger mechanism \cite{schwinger,schwinger1}, according to which the probability distribution of the emission of charged particles with mass $m_{\nu}$ and transverse momentum ${p_t}_\nu$ is proportional to  ${g_{\nu}  \exp{ \left( -\frac{\pi ({p_{t}}_\nu^2 + m_{\nu}^2) } {t}\right)}}.$ Here $t$ is a string tension, and $g_\nu$ is a weighting factor, responsible for cascade fragmentation like the $\rho$ meson decay into several pions. In the present model, the string tension is modified according to usual string fusion prescription: $t=t_{0}\sqrt{\eta}$, where $\eta$ is the number of overlapped strings. The values of ${p_t}_\nu$ and $g_\nu$ are listed in Table \ref{tab1}.

\begin{center}
\begin{table}[h]
\caption{\label{tab1}Parameters of particles differentiation.}
%\footnotesize\rm
\centering
\begin{tabular}{@{}*{7}{l}}
\br
 $\nu$ \hspace{1.0cm} & Particle species  \hspace{1.7cm} &  $m_\nu, \mathrm{GeV}$ \hspace{1.7cm} &   $g_\nu$ \hspace{1.7cm} \\ \mr
0 & Pions from $\rho$ decays & 0.775 & 3 \\ 
1 & Direct pions & 0.135 & 1 \\ 
2 & Kaons & 0.494 & 1 \\ 
3 & Protons & 0.938 & 1 \\ 
\br
\end{tabular}
\end{table}
\end{center}

\section{Results}
\begin{figure}[hb]
\begin{center}
\begin{minipage}{14pc}
\includegraphics[width=14pc]{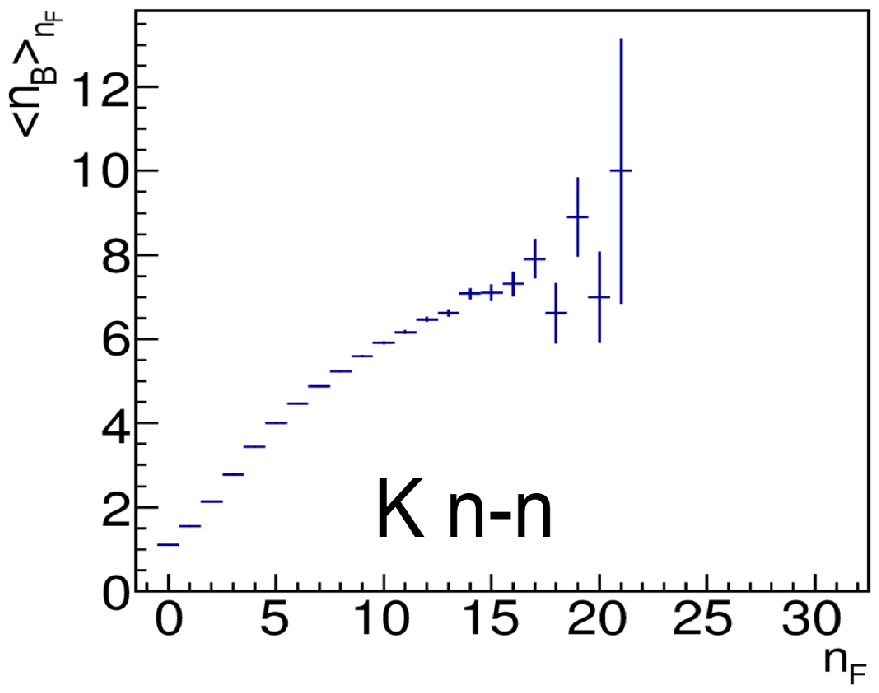}
\vspace{0.2cm}
\end{minipage}\hspace{2pc}%
\begin{minipage}{15pc}
\includegraphics[width=15pc]{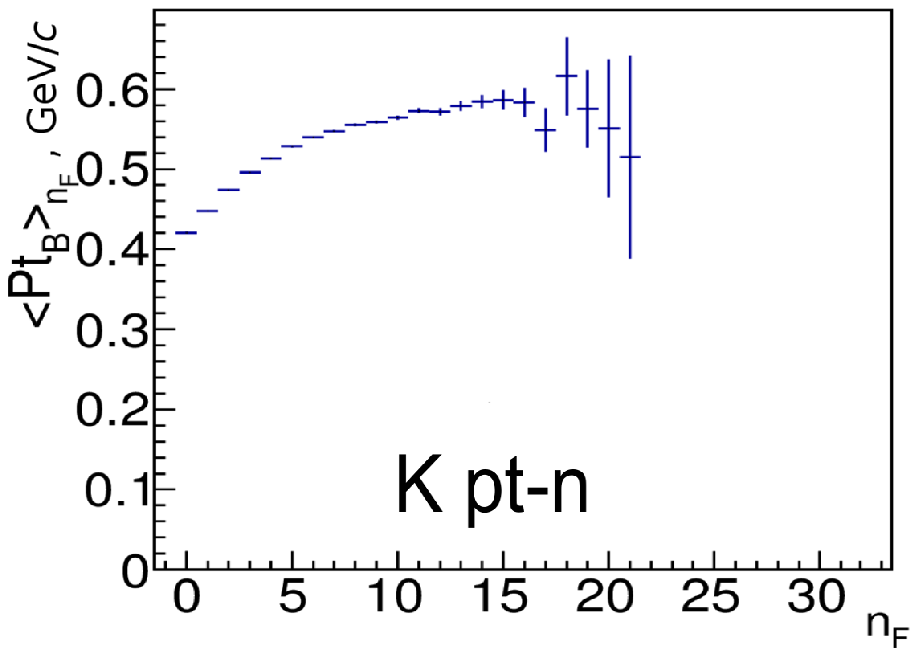}
\vspace{0.2cm}
\end{minipage} 
\includegraphics[width=14.2pc]{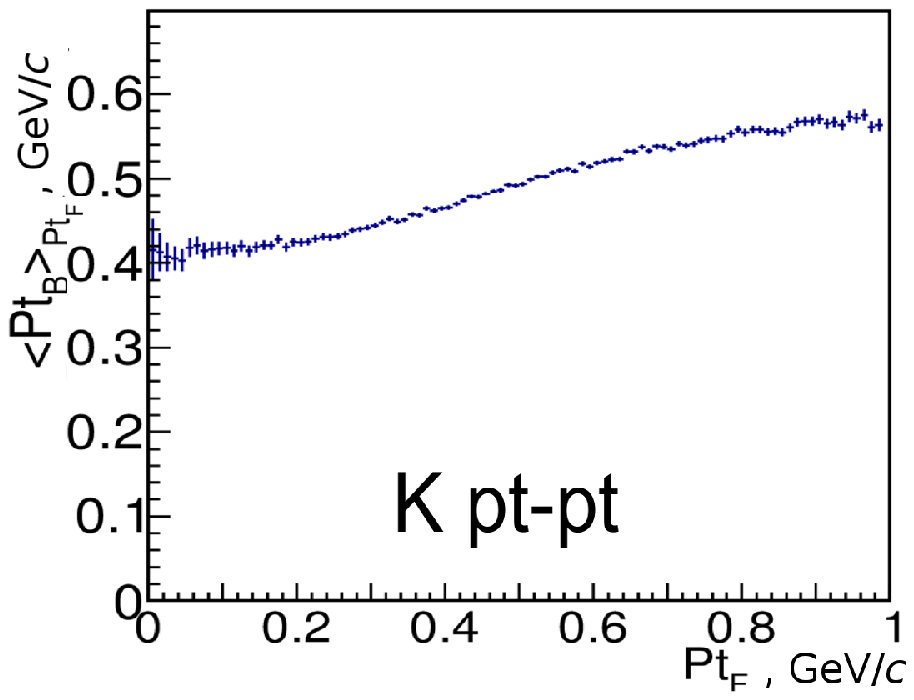}\hspace{4.2pc}%
\begin{minipage}[b]{13pc}\caption{\label{Figure1}Correlation functions for kaons, produced in $pp$ collisions at $\sqrt{s}=$~7~TeV in rapidity windows (-2.5, -0.5), (0.5, 2.5): $n-n$ (left), $p_t-n$ (right), $p_t-p_t$ (bottom), calculated in Monte Carlo model with string fusion.\vspace{0.65cm}}
\end{minipage}
\end{center}
\end{figure}

In Figure \ref{Figure1} the $n-n$, $p_t-n$ and $p_t-p_t$ correlation functions for kaons are shown. All three correlation functions have non-linear behavior with a tendency of saturation at high multiplicities and transverse momenta. These features are similar to those of correlation functions of non-identified hadrons  \cite{MC_model2}; however, they differ quantitatively.

\begin{figure}[ht]
\begin{center}
\vspace{-1cm}
\includegraphics[width=28.5pc]{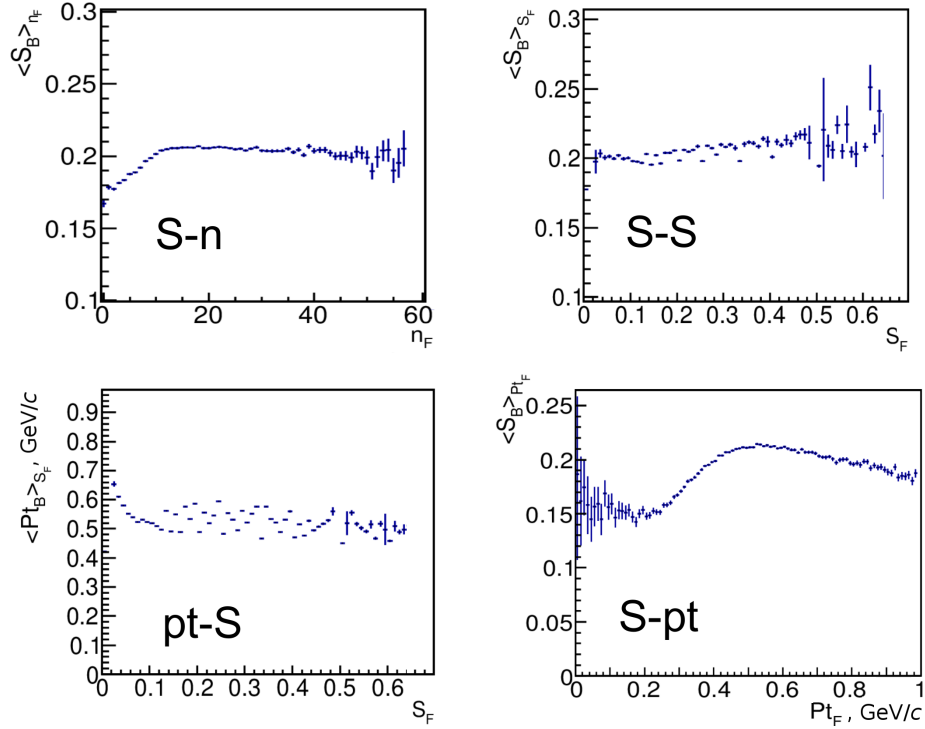}
\caption{\label{Figure2}Correlation functions in $pp$ collisions at $\sqrt{s}=$~7~TeV in rapidity windows (-2.5, -0.5), (0.5, 2.5): $S-n$ (top left), $S-S$ (top right), $p_t-S$ (bottom left), $S-p_t$ (bottom right), calculated in Monte Carlo model with string fusion.}
\end{center}
\vspace{-0.5cm}

\end{figure}

In Figure \ref{Figure2} the results for new types of correlation functions are shown. They could be calculated only by taking into account the particle differentiation. One can see that the fraction of the strange particles ($S$) in the backward window behaves non-monotonously with charged multiplicity in the forward window: it increases at small multiplicities, then saturates and even decreases at high multiplicities. Similar behaviour is obtained for strangeness -- transverse momentum ($S-p_t$) correlation function. This could be related to the growth of the non-strange resonance fraction at high string density, which after a strong decay produce more pions than strange particles.  Comparison of $p_t-S$ and $S-p_t$ correlations demonstrates that the use of the transverse momentum in the forward window for event classification provides more pronounced correlation than the strangeness fraction. 

The results also show non-zero strangeness-strangeness correlation, although the slope is rather smaller in $pp$ collisions. We should note that these dependencies could be obtained only when the model includes string fusion, without the fusion the correlations would be zero. Therefore, these correlations could serve as an indicator of string fusion when comparing to experimental data.

It is important, that the strangeness fraction, as well as mean event transverse momentum is an intensive variable. Hence, the $S-S$, $p_t-S$ correlation coefficients are expected to be robust against the peculiarities of the event selection in the collisions with nuclei, as it was established for $p_t-p_t$ correlations \cite{PtPtEPJWOC}.

\section{Conclusions}

Long-range correlations between observables in separated rapidity window are calculated in the framework of the string fusion model, taking into account finite rapidity length of the strings, hardness of the process and particle differentiation according to Schwinger mechanism. The $n-n$, $p_t-n$ and $p_t-p_t$ correlation functions for strange particles have been studied. Four new types of correlation functions are introduced by an accounting of the mean event strangeness. The predictions for $pp$ interactions at 7~TeV are obtained. 

\ack

The authors acknowledge Saint-Petersburg State University for a research grant 11.38.242.2015. V.K. acknowledges the support of Special SPbSU Rector's Scholarship, G-RISC stipend and Dynasty Foundation Scholarship.

\end{document}